\documentclass[debug]{rmaa}

\title{Planetary nebulae and the chemical evolution of the Magellanic Clouds} 

\author{
  W. J. Maciel,\altaffilmark{1} 
  R. D. D. Costa,\altaffilmark{1}
  and T. E. P. Idiart\altaffilmark{1}}

\altaffiltext{1}{Instituto de Astronomia, Geof\'\i sica e Ci\^encias
                 Atmosf\'ericas, Universidade de S\~ao Paulo, Brazil.}

\shortauthor{Maciel, Costa, \& Idiart}
\shorttitle{PN and the chemical evolution of the MC}

\fulladdresses{
\item W. J. Maciel, R. D. D. Costa and T. E. P. Idiart: 
    Instituto de Astronomia, Geof\'\i sica e Ci\^encias
    Atmosf\'ericas, Universidade de S\~ao Paulo - 
    Rua do Mat\~ao 1226, CEP 05508-900, S\~ao Paulo SP, Brazil
    (maciel@astro.iag.usp.br, roberto@astro.iag.usp.br, 
     thais@astro.iag.usp.br.)}

\listofauthors{W. J. Maciel, R. D. D. Costa, \& T. E. P. Idiart}
\indexauthor{Maciel, W. J.}
\indexauthor{Costa, R. D. D.}
\indexauthor{Idiart, T. E. P.}

\abstract{The determination of accurate chemical abundances of planetary nebulae (PN) 
in different galaxies allows us to obtain important constraints of chemical evolution 
models for these systems. We have a  long term program to derive abundances in the 
galaxies of the Local Group, particularly the Large and Small Magellanic Clouds. In this 
work, we present our new results on these objects and discuss their implications in view 
of recent abundance determinations the literature. In particular, we obtain distance-independent 
correlations involving He, N, O, Ne, S, and Ar, and compare the results with data 
from our own Galaxy and other galaxies in the Local Group. As a result of our 
observational program, we have a large database of PN in the Galaxy and the Magellanic
Clouds, so that we can obtain reliable constraints to the nucleosynthesis processes in 
the progenitor stars in galaxies of different metallicities.}

\resumen{ }

\addkeyword{ISM: planetary nebulae: general}
\addkeyword{galaxies: Magellanic Clouds}
\addkeyword{galaxies: abundances}

\begin{document}
% Typeset article header
\maketitle

%---------------------------------------------------------------------------------------------
\section{Introduction}
\label{section1}
%---------------------------------------------------------------------------------------------

The study of the chemical evolution of the galaxies in the Local Group, particularly
the Milky Way and the Magellanic Clouds, can be significantly improved by the consideration
of the chemical abundances of planetary nebulae (PN) (see for example Maciel et al. 
\citeyear{maciel2006a}, Richer \& McCall \citeyear{richer}, Buzzoni et al. \citeyear{buzzoni}, 
and Ciardullo \citeyear{ciardullo}). These objects are produced by low and intermediate 
mass stars, with main sequence masses roughly between 0.8 and 8$\,M_\odot$, and present
a reasonably large age and metallicity spread. As a conclusion, they provide important 
constraints to the chemical evolution models applied to these systems, and can also be 
used to test nucleosynthetic processes in the PN progenitor stars. In particular, the PN
abundances in the nearby Magellanic Clouds can be derived with a high acuracy, comparable
to the objects in the Milky Way, so that they can be especially useful in the study
of the chemical evolution of these galaxies. In this work, we present some recent results 
on the determination of chemical abundances from PN in the Large and Small Magellanic Clouds
derived by our group, and compare these results with recent data from our own Galaxy and 
other galaxies in the Local Group. We also take advantage of the inclusion of similar 
determinations from the recent literature, so that the  database of PN in the Magellanic 
Clouds is considerably increased, allowing  a better determination of observational 
constraints of the nucleosynthetic processes ocurring in the progenitor stars. Preliminary 
results of this work have been presented by Maciel, Costa \& Idiart (\citeyear{maciel2006a},
\citeyear{keele}).

%---------------------------------------------------------------------------------------------
\section{The Sample}
\label{section2}
%---------------------------------------------------------------------------------------------

We have considered a sample of PN both in the LMC and SMC on the basis of observations 
secured at the 1.6m LNA telescope located in southeast Brazil and the ESO 1.5m telescope
in La Silla, Chile. Details of the observations and the resulting abundances can be found 
in the following references: de Freitas Pacheco et al. (\citeyear{pacheco1}, 
\citeyear{pacheco2}), Costa et al. (\citeyear{costa1}), and Idiart et al.
(\citeyear{idiart}). In these papers, abundances of He, N, O, S, Ne and Ar have been 
determined for 23 nebulae in the LMC and 46 objects in the SMC. The abundances presented
in Idiart et al. (\citeyear{idiart}) were based on average fluxes obtained by taking into
account some recent results from the literature, so that there may be some differences
compared with our originally derived values. For details the reader is referred to the
discussion in that paper. 

In order to increase the PN database in the Magellanic Clouds, we have also taken into account 
the samples by Stasi\'nska et al. (\citeyear{stasinska}), which included abundances of He, N, O, 
and Ne for 61 nebulae in the SMC and 139 objects in the LMC, and Leisy \& Dennefeld  
(\citeyear{leisy}), containing 37 objects in the SMC and 120 nebulae in the LMC. In Stasi\'nska 
et al. (\citeyear{stasinska}), a collection was obtained of photometric and spectroscopic data of 
PN in five different galaxies, including the Magellanic Clouds. Although the original sources 
of the data are rather heterogeneous, the plasma diagnostics and determination of the chemical 
abundances were processed in the same way, so that the degree of homogeneity of the data 
was considerably increased. The Leisy \& Dennefeld (\citeyear{leisy}) sample is a more 
homogeneous one, in which a larger fraction of the observations were made by the authors 
themselves, and all abundances were re-derived in an homogeneous way, as in Stasi\'nska et 
al. (\citeyear{stasinska}). As we will show in the next section, the similarity of the methods 
in the abundance determinations warrants comparable abundances, so that a larger sample was 
obtained.

%---------------------------------------------------------------------------------------------
\section{Results and discussion}
\label{section3}
%---------------------------------------------------------------------------------------------

\subsection{Average abundances}

Average abundances of all elements in the SMC and LMC according to the three samples considered 
are shown in Table~1. Helium abundances are given as He/H by number as usual, while for the
heavier elements the quantity given is $\epsilon({\rm X}) = \log {\rm X/H} + 12$.
Although the samples considered here are probably the largest ones with carefullly
derived abundances in the Magellanic Clouds, they cannot be considered as complete.
The total number of PN in these systems is not known, but recent estimates point to
about 130 and 980 objects for the SMC and LMC, respectively (cf. Jacoby \citeyear{jacoby}
and Shaw \citeyear{shaw}). Therefore, incompleteness effects may still affect the results 
presented in this paper.

The He abundances show a good agreement in all samples within the average uncertainties.
The IAG and Leisy samples show a slightly higer He abundance in the LMC compared to
the SMC, but the differences between these objects are in all cases smaller than the
estimated uncertainties. 

The O/H abundances, which are in general the best determined
of all heavy elements considered here, also show a good agreement among the samples.
Moreover, in all cases the LMC is richer than the SMC, as expected, and the average
metallicity difference is in the range 0.3 to 0.5 dex, which is consistent with the
metallicities given by Stanghellini (\citeyear{stanghellini08}), namely $Z = 0.004$
and $Z = 0.008$ for the SMC and LMC, respectively.

The Ar/H and Ne/H ratios show a similar behaviour as O/H, noticing that the IAG data
do not include Ne abundances for the LMC, and that Stasi\'nska et al. 
(\citeyear{stasinska}) do not list Ar/H abundances for both galaxies. The sulfur 
abundances seem to be less reliable, as can be seen from the large standard deviations 
obtained in the IAG and Leisy \& Dennefeld samples. Moreover, the estimated average S/H 
ratio in the SMC is slightly larger than in the LMC according to the IAG data, contrary 
to our expectations, while in the Leisy \& Dennefeld sample the LMC abundance is larger 
by only 0.27 dex in comparison with the SMC. In fact, these characteristics of the sulfur 
abundances in Magellanic Cloud PN can be observed in previous analyses, as for example 
in the summary by Kwok (\citeyear{kwok}, Table 19.1, p. 202), where the S/H ratios in the
SMC and LMC are indistinguishable within the given uncertainties. Clearly, the determination 
of S/H abundances in the Magellanic Cloud PN - and galactic nebulae as well - is apparently 
affected by some additional effects, as compared to the previous elements. In the 
following we will give further evidences on the problem of sulfur abundances in 
planetary nebulae.

From Table~1, it can be seen that the nitrogen abundances also follow the same pattern as O/H, 
Ar/H, and Ne/H, even though the N/H ratio is affected by the dredge-up episodes occuring in the PN
progenitor stars. This is further discussed in Section~5, but from the results shown in
the last column of Table~1, it is suggested that the average nitrogen contamination from the
PN progenitor stars is small. Average N/H abundances of of Magellanic Cloud PN are given
by Stanghellini (\citeyear{stanghellini08}), where an effort was made to take into account
objects of different morphologies. The average N/H abundances in the whole sample
are $1.48 \times 10^{-4}$ and $0.29 \times 10^{-4}$ for the LMC and SMC, respectively,
which correspond to $\epsilon({\rm N}) = 8.17$ and 7.46, in the notation of Table~1.
These results correctly indicate that the LMC is richer in N than the SMC, as also reported
in Table~1, and the absolute value of the SMC abundances given by Stanghellini 
(\citeyear{stanghellini08}) is very similar to the results of the 3 samples considered here,
but the average abundance for the LMC nebulae is much higher than our results. 
In fact, the N/H abundances of the LMC given in Stanghellini (\citeyear{stanghellini08}) 
are close to the Milky Way values, which is paradoxical, as the LMC has a much lower
metallcity than the Galaxy. Part of the discrepancy may be caused by the fact that the
sample used in that paper includes a larger proportion of
bipolar nebulae, which are ejected by higher mass progenitor stars, which produce a larger
nitrogen contamination than the lower mass objects. This problem needs further clarification.

\begin{table*}
\small
\begin{changemargin}{-2cm}{-2cm}
\caption[]{Average abundances of PN in the Magellanic Clouds.}
\label{table1}
\begin{flushleft}
\begin{tabular}{lcccccc}
\noalign{\smallskip}
\hline\noalign{\smallskip}
 & He &  O &  S & Ar & N & Ne \\
\noalign{\smallskip}
\hline\noalign{\smallskip}
IAG/USP & & & & & & \\
SMC & $0.097\pm 0.035$ & $7.89\pm 0.44$ & $6.98\pm 0.58$ & $5.59\pm 0.36$ & $7.35\pm 0.49$ & $7.14\pm 0.42$ \\
LMC & $0.119\pm 0.023$ & $8.40\pm 0.20$ & $6.72\pm 0.31$ & $6.01\pm 0.25$ & $7.69\pm 0.50$ & --- \\
    & & & & & & \\
Stasi\'nska & & & & &  & \\
SMC & $0.094\pm 0.025$ & $7.74\pm 0.50$ & --- & --- & $7.46\pm 0.37$ & $7.10\pm 0.40$ \\
LMC & $0.090\pm 0.032$ & $8.10\pm 0.31$ & --- & --- & $7.76\pm 0.45$ & $7.44\pm 0.41$ \\
    & & & & & & \\
Leisy & & & & &  & \\
SMC & $0.093\pm 0.025$ & $8.01\pm 0.29$ & $6.86\pm 0.67$ & $5.57\pm 0.27$ & $7.39\pm 0.47$ & $7.14\pm 0.36$ \\
LMC & $0.105\pm 0.035$ & $8.26\pm 0.35$ & $7.13\pm 0.67$ & $5.99\pm 0.26$ & $7.77\pm 0.57$ & $7.46\pm 0.48$ \\
    & & & & & & \\
Orion & $0.098\pm 0.004$ & $8.55\pm 0.07$ & $7.02\pm 0.10$ & $6.52\pm 0.18$ & $7.78\pm 0.12$ & $7.82\pm 0.16$ \\
Sun & $0.092\pm 0.009$ & $8.80\pm 0.11$ & $7.26\pm 0.08$ & $6.48\pm 0.11$ & $7.97\pm 0.07$ & $8.08\pm 0.01$ \\
30 Dor  & $0.087\pm 0.001$ & $8.33\pm 0.02$ & $6.84\pm 0.10$ & $6.09\pm 0.10$ & $7.05\pm 0.08$ & $7.65\pm 0.06$ \\
NGC 346 & ---              & $8.15$         & $6.40$         & $5.82$         & $6.81$         & $7.32$         \\
\noalign{\smallskip}
\hline
\end{tabular}
\end{flushleft}
\end{changemargin}
\end{table*}

\subsection{Abundances of individual nebulae}

% para incluir figura girando 90 graus
% \includegraphics[width=\textwidth]{fig1.eps}
% para incluir figura girando 90 graus
%   \includegraphics[angle=-90,width=14.0cm]{fig1.eps}
%
%oooooooooooooooooooooooooooooooooooooooooooooooooooooooooooooooooooooooooooooooooooooooooooooo
   \begin{figure}
   \centering
   \includegraphics[angle=-90]{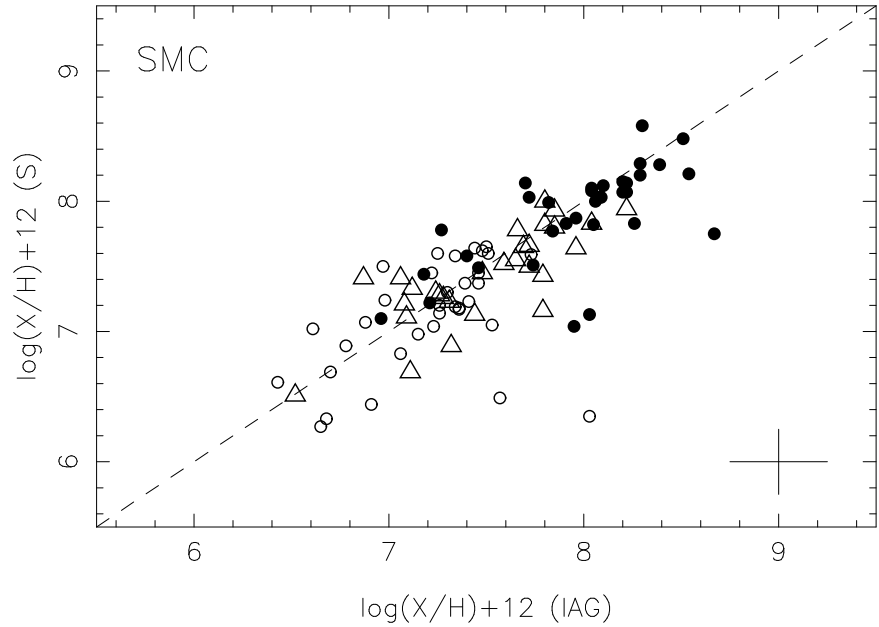}
      \caption{Abundances of O/H (solid dots), N/H (empty triangles)
      and Ne/H (empty circles) from the sample by Stasi\'nska et al.
      (\citeyear{stasinska}) as a function of data from the IAG/USP
      group for the SMC.}
   \label{fig1}
   \end{figure}

%oooooooooooooooooooooooooooooooooooooooooooooooooooooooooooooooooooooooooooooooooooooooooooooo
   \begin{figure}
   \centering
   \includegraphics[angle=-90]{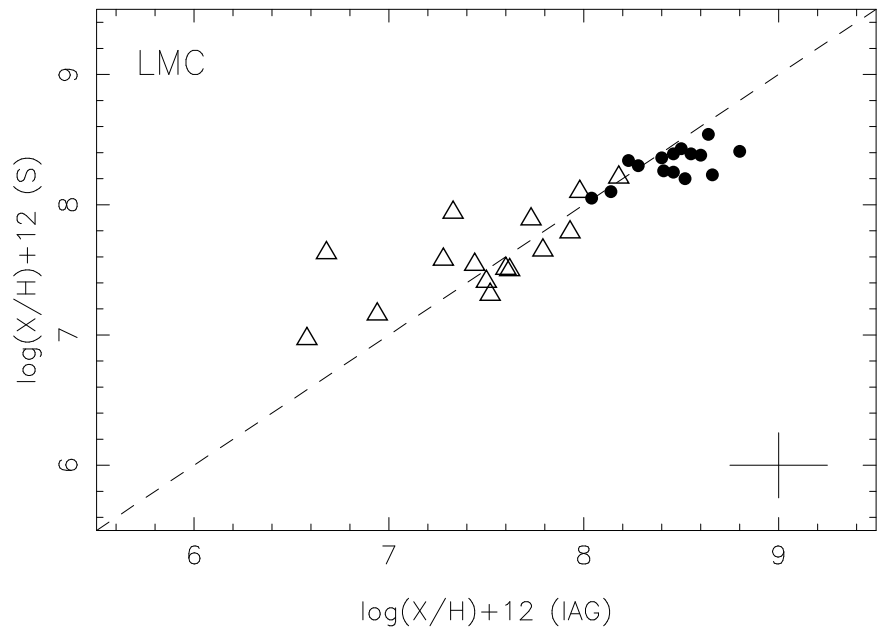}
      \caption{The same as Fig.~1 for the LMC. No Ne/H data is available
      in the IAG sample for this galaxy.}
   \label{fig2}
   \end{figure}

%oooooooooooooooooooooooooooooooooooooooooooooooooooooooooooooooooooooooooooooooooooooooooooooo
   \begin{figure}
   \centering
   \includegraphics[angle=-90]{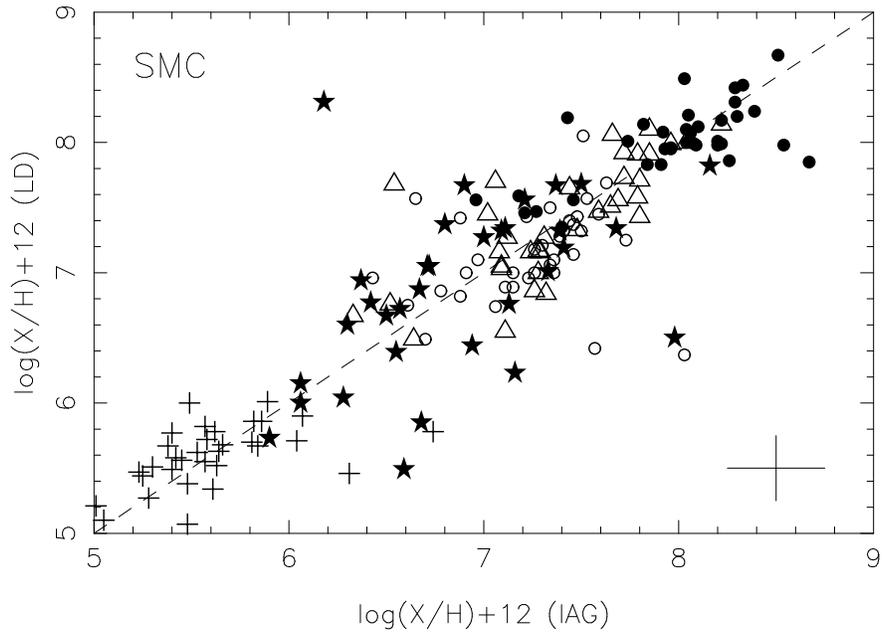}
      \caption{Abundances of O/H (solid dots), S/H (stars), Ar/H (crosses),
      N/H (empty triangles), and Ne/H (empty circles) from the sample 
      by Leisy \& Dennefeld (\citeyear{leisy}) as a function of data from 
      the IAG/USP group for the SMC.}
   \label{fig3}
   \end{figure}

%oooooooooooooooooooooooooooooooooooooooooooooooooooooooooooooooooooooooooooooooooooooooooooooo
   \begin{figure}
   \centering
   \includegraphics[angle=-90]{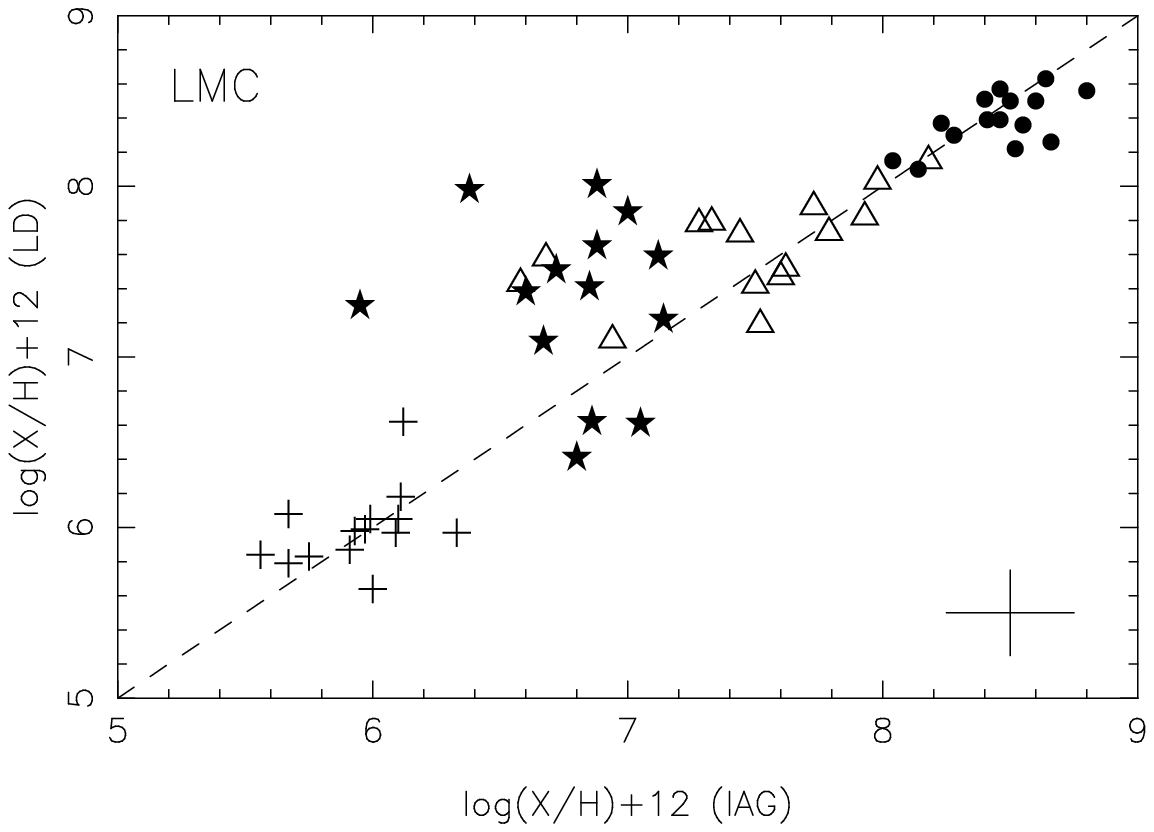}
      \caption{The same as Fig.~3 for the LMC. No Ne/H data is available
      in the IAG sample.}
   \label{fig4}
   \end{figure}

In order to illustrate the internal agreement of the three PN samples considered in this work, 
we present in Figs.~1 and 2 the abundances of O/H (solid dots), N/H (empty triangles) and 
Ne/H  (empty circles) as derived by Stasi\'nska et al. (\citeyear{stasinska}) as a function 
of the IAG/USP data, for the SMC and LMC nebulae, respectively. The same comparisons are shown 
in Figs.~3 and 4 taking into account the data by Leisy \& Dennefeld (\citeyear{leisy}), in 
which case we also include abundances of S/H (stars) and Ar/H data (crosses). An average
error bar is included at the lower right corner of the figures. The agreement 
of both samples with our own data is generally very good, within the average uncertainties of 
the abundance data, which are about 0.1 to 0.2 dex for the best derived abundances, and of 
0.2 to 0.3 for the less accurate element ratios. Some scatter is to be expected, especially 
taking into account that the abundances of several nebulae are flagged as uncertain (:) by
Leisy \& Dennefeld (\citeyear{leisy}). The main discrepancies between the IAG data and 
the results by Stasi\'nska et al. (\citeyear{stasinska}) occur for a few objects in the SMC,
for which our O/H and Ne/H abundances differ by an amount larger than the average uncertainty
(cf. Fig.~1), while for the LMC a small group of nebulae have higher N/H abundances as derived
by Stasi\'nska et al. (\citeyear{stasinska})  (cf. Fig.~2). Concerning the Leisy \& Dennefeld 
(\citeyear{leisy}) sample, the main discrepancies are restricted to some S/H data, as can be
seen from Figs.~3 and 4. The origin of these discrepancies is not clear, but it should be
stressed that the vast majority of the objects in common in the 3 samples have similar results,
as illustrated in Figs.~1 to 4.

\subsection{Metallicity differences: the Galaxy and the Magellanic Clouds}

The PN abundances of the heavy elements O, S, Ne, and Ar as given in Table~1 are expected
to reflect the average metallicities of the Magellanic Clouds, which are a few dex
lower than in the Milky Way, as these elements are not produced by the PN progenitor
stars. This can be confirmed by comparing the PN abundances with the abundances in the
Orion Nebula, which can be taken as representative of the present heavy element abundances
in the Galaxy. From the compilation by Stasi\'nska (\citeyear{stasinska2004}), we obtain
the abundances given at the end of Table~1. For comparison purposes, the average solar
abundances from the same source are also included. For the Orion Nebula and the Sun
the uncertainties given are not the intrinsic uncertainties of the data, but the dispersion
of the measurements in the recent literature as considered by Stasi\'nska 
(\citeyear{stasinska2004}). It can be seen that the Orion Nebula abundances are higher than
the Magellanic Clouds PN by about 0.2 to 0.5 dex for the LMC and 0.5 to 0.8  dex for the
SMC in the case of oxygen.  The average for the Orion Nebula, $\epsilon_{ON}({\rm O})  
\simeq 8.55$, is also essentially the same as in the galactic PN,
$\epsilon_{PN}({\rm O}) \simeq 8.65$ (cf. Maciel et al. \citeyear{maciel2006a}).
For S, Ne, and Ar a similar comparison is obtained, although the S abundance in the LMC is 
actually somewhat higher than in the Orion Nebula according to the data by Leisy \& Dennefeld 
(\citeyear{leisy}). The difference in the abundances is also smaller in the case of nitrogen,  
which is a clear evidence of the N enhancement in the PN progenitor stars. 

\subsection{The metallicity distribution}

%oooooooooooooooooooooooooooooooooooooooooooooooooooooooooooooooooooooooooooooooooooooooooooooo
   \begin{figure}
   \centering
   \includegraphics[width=12.0cm]{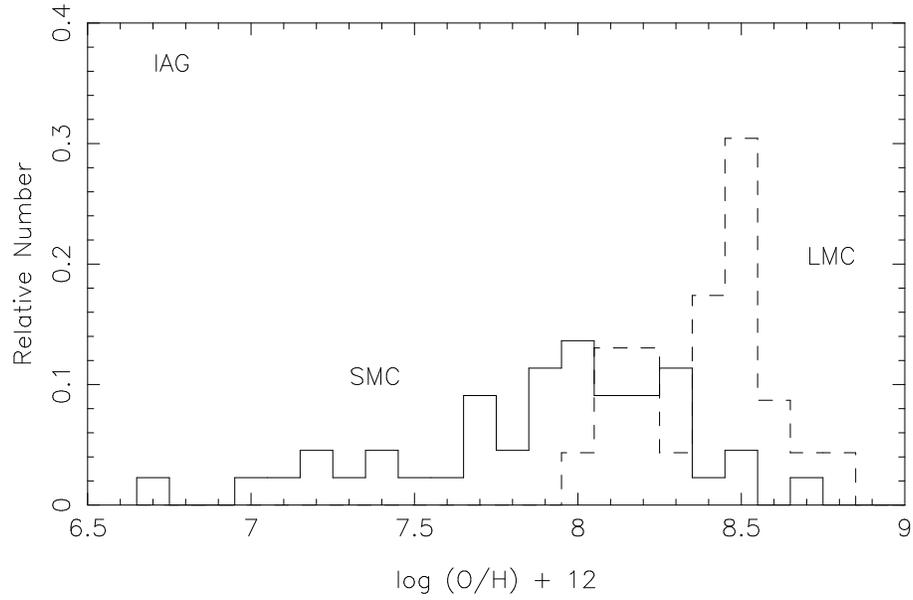}
      \caption{The O/H abundance distribution of the Magellanic Clouds
      from the IAG/USP data.}
   \label{fig5}
   \end{figure}

%oooooooooooooooooooooooooooooooooooooooooooooooooooooooooooooooooooooooooooooooooooooooooooooo
   \begin{figure}
   \centering
   \includegraphics[width=12.0cm]{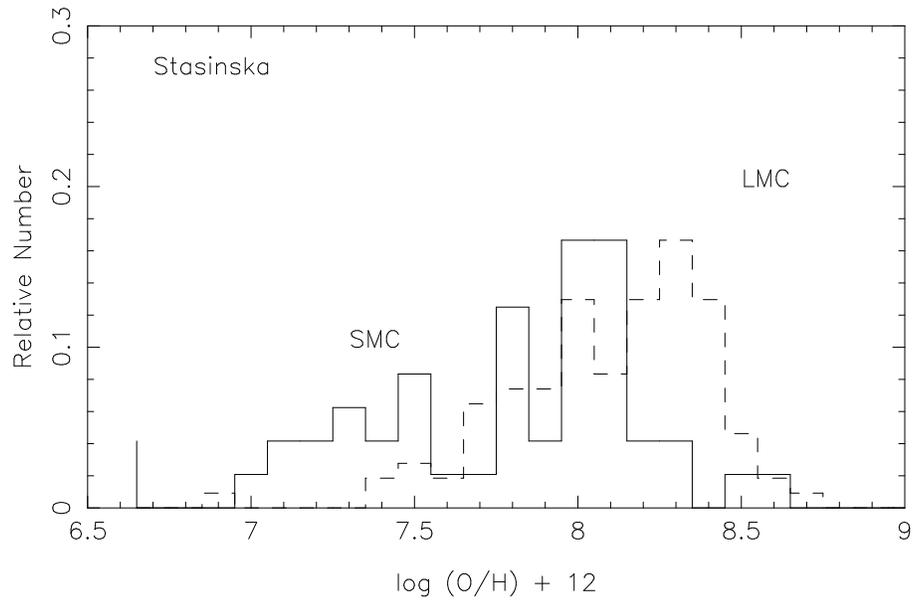}
      \caption{The same as Fig.~5 for the data by Stasi\'nska et al.
       (\citeyear{stasinska}).}
   \label{fig6}
   \end{figure}

%oooooooooooooooooooooooooooooooooooooooooooooooooooooooooooooooooooooooooooooooooooooooooooooo
   \begin{figure}
   \centering
   \includegraphics[width=12.0cm]{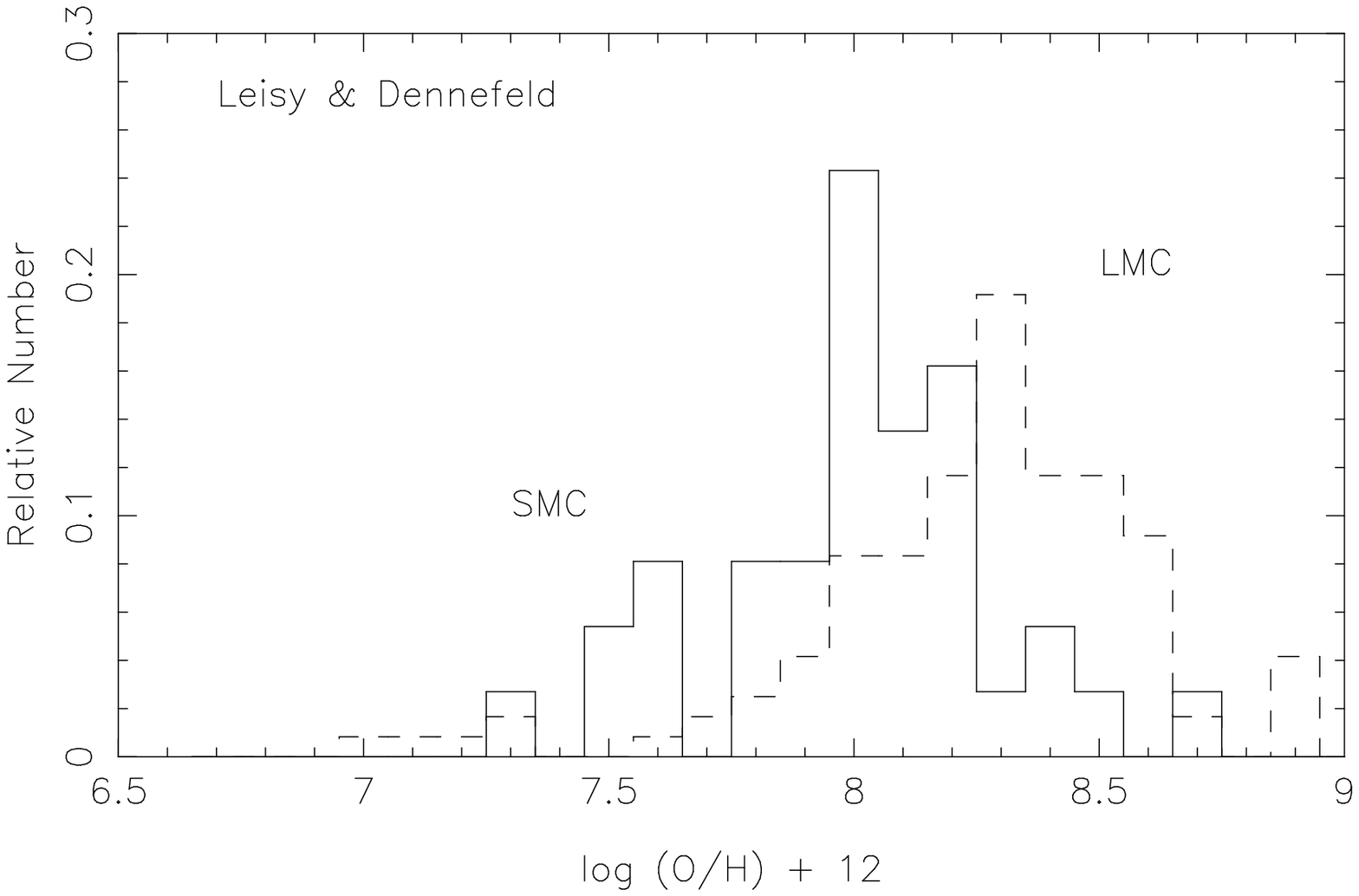}
      \caption{The same as Fig.~5 for the data by Leisy \& Dennefeld
       (\citeyear{leisy}).}
   \label{fig7}
   \end{figure}

The available data on PN in the Magellanic Clouds can be used to infer the metallicity 
distribution in these systems, on the basis of the derived abundances of O, S, Ne, and 
Ar. A comparison of the distributions in different systems can be used to infer their 
average  metallicities, with consequences on the star formation rates. As an example, 
Figs.~5. 6 and 7  show the O/H  distribution in the Magellanic Clouds according to the
three samples considered in this work. The metallicity difference between the
SMC and the LMC can be clearly observed in all samples, amounting to about 0.4 to 0.5 dex
in average. The difference is especially well defined in our sample, as shown in Fig.~5.
In a comparison with the Milky Way, the galactic disk nebulae extend to a higher metallicity, 
up to $\epsilon({\rm O}) \simeq 9.2$, while the LMC objects reach $\epsilon({\rm O}) \simeq 8.8$ 
and the lowest metallicities in the SMC are about $\epsilon({\rm O}) \simeq 7.0$. Concerning 
the remaining elements that are not affected by the evolution of the PN progenitor 
stars, the Ar/H abundance distribution has a similar pattern, while the S/H data is less
clear, as already mentioned. We will discuss this element in more detail in the next section. 
For Ne/H we have no IAG data for the LMC, but the larger Leisy \& Dennefeld 
(\citeyear{leisy}) sample clearly confirms the 0.4 to 0.5 dex difference between the LMC 
and the SMC.

The metallicity distribution of the PN as shown in Figs.~5, 6, and 7 can also be
compared with galactic data, both for disk and bulge PN. Cuisinier et al. 
(\citeyear{cuisinier}) considered a sample of 30 bulge nebulae and a compilation 
containing about 200 disk PN, and concluded that both O/H distributions are similar,
peaking around 8.7--8.8 dex, and extending form $\epsilon({\rm O}) \simeq 8.0$\ to
$\epsilon({\rm O}) \simeq 9.2$. More recently, Escudero et al. (\citeyear{escudero}) 
obtained a similar distribution using a bulge sample twice as large, which extended
to about 7.5 dex (see also Costa et al. \citeyear{costa3}). According to Figs.~5--7, 
the O/H distributions are displaced relative to the Milky Way by approximately 0.4 and 
0.7 dex towards shorter metallicities for the LMC and SMC, respectively, in good agreement 
with the results discussed in Section~4.1.

\subsection{Abundance correlations: elements not produced by the progenitor stars}

Photoionized nebulae, comprising both PN and HII regions, are extremely useful to study 
chemical abundances in different systems. While HII regions reflect the present chemical 
composition of star-forming systems, PN is helpful to trace the time evolution of the 
abundances, especially when an effort is made to establish their age distribution 
(see for example Maciel et al. \citeyear{maciel2006b}). The elements S, Ar and Ne are 
probably not produced by the PN progenitor stars, as they are manufactured in the late 
evolutionary stages of massive stars. Therefore, S, Ar, and Ne abundances as measured in 
PN should reflect the interstellar composition at the time the progenitor stars were formed. 
Since in the interstellar medium of star-forming galaxies such as the Magellanic Clouds 
the production of O and Ne is believed to be dominated by type II supernovae, we may conclude 
that the original O and Ne abundances are not significantly modified by the stellar progenitors 
of bright PN. 

The variation of the ratios S/H, Ar/H and Ne/H with O/H usually show a good positive 
correlation for all studied systems in the Local Group, with similar slopes close to unity. 
The main differences lie in the average metallicity of the different galaxies, which can be 
inferred from the observed metallicity range, as we have seen in the previous section. 

Fig.~8 shows the Ne/H ratio as a function of O/H for the SMC, while Fig.~9 corresponds to the 
LMC. In these figures we include the combined samples mentioned in Section~2 as follows:
IAG/USP data (filled circles), Stasi\'nska et al. (\citeyear{stasinska}) (empty circles), and
Leisy \& Dennefeld (\citeyear{leisy}) (crosses). Average error bars are included at the lower 
right corner of the figures. It can be seen that the correlation is very
good, with a slope in the range 0.8--0.9 in both cases. The Ne/H $\times$\ O/H relation
is probably the best example provided by PN regarding the nucleosynthesis in massive stars.
This correlation is very well defined, as shown in Figs.~8 and 9, and is essentially the
same as derived from HII regions in different star forming galaxies of the Local Group,
including the Milky Way, and in emission line galaxies as well, as clearly shown by 
Richer \& McCall (\citeyear{richer}) and Richer (\citeyear{richer2}, see also Henry et al. 
\citeyear{henry06}).

The Ar/H data shows a similar correlation with O/H, as can be seen from Figs.~10 and 11, 
but the correlation is poorer, which may be due to the fact that the samples are smaller, 
since Stasi\'nska et al. (\citeyear{stasinska}) do not present argon data. Again, the main 
discrepancy lies in the S/H data, as can be seen from Figs.~12 and 13. Although most objects 
define a positive correlation, which is especially true for the LMC, the dispersion is much 
larger in the S/H data compared to the previous elements, again suggesting that a problem 
remains in the interpretation of the S/H abundances in planetary nebulae. In particular, 
both the IAG/USP and Leisy \& Dennefeld data suggest a scattering diagram on the S/H $\times$\ 
O/H plane for the SMC, with an average abundance around $\epsilon({\rm S/H}) = \log ({\rm S/H}) 
+ 12 \simeq 7.0$. A weaker correlation involving sulfur is to be expected, since the
diagnostic lines for this element are weaker than e.g.  for oxygen or neon. However, 
the real situation may be more complex, so that a more detailed discussion is appropriate.

%oooooooooooooooooooooooooooooooooooooooooooooooooooooooooooooooooooooooooooooooooooooooooooooo
   \begin{figure}
   \centering
   \includegraphics[angle = -90]{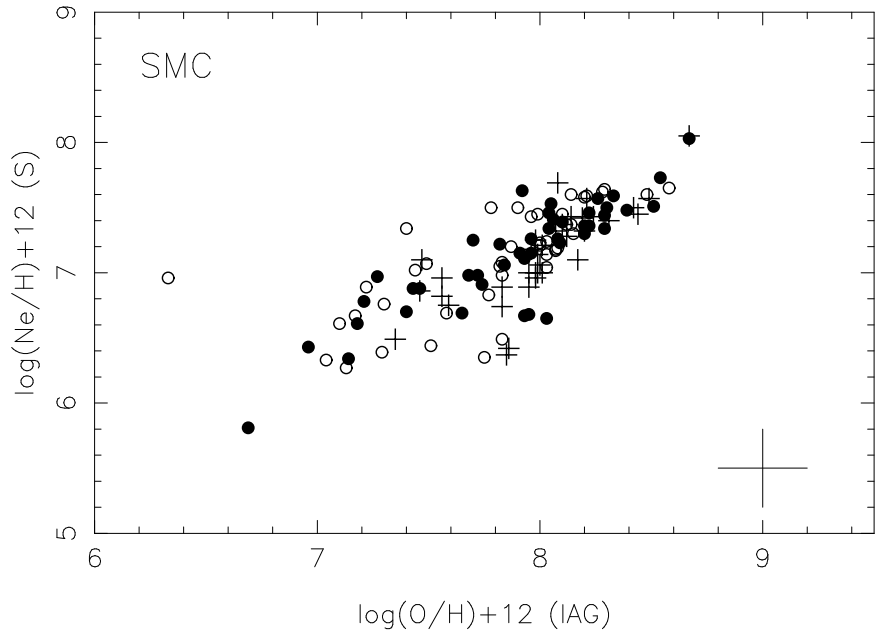}
      \caption{Distance-independent correlation of Ne/H $\times$ O/H for
      the SMC. Filled circles: IAG/USP data; empty circles: Stasi\'nska
      et al. (\citeyear{stasinska}); crosses: Leisy \& Dennefeld 
      \citeyear{leisy}).}
   \label{fig8}
   \end{figure}
%oooooooooooooooooooooooooooooooooooooooooooooooooooooooooooooooooooooooooooooooooooooooooooooo
   \begin{figure}
   \centering
   \includegraphics[angle = -90]{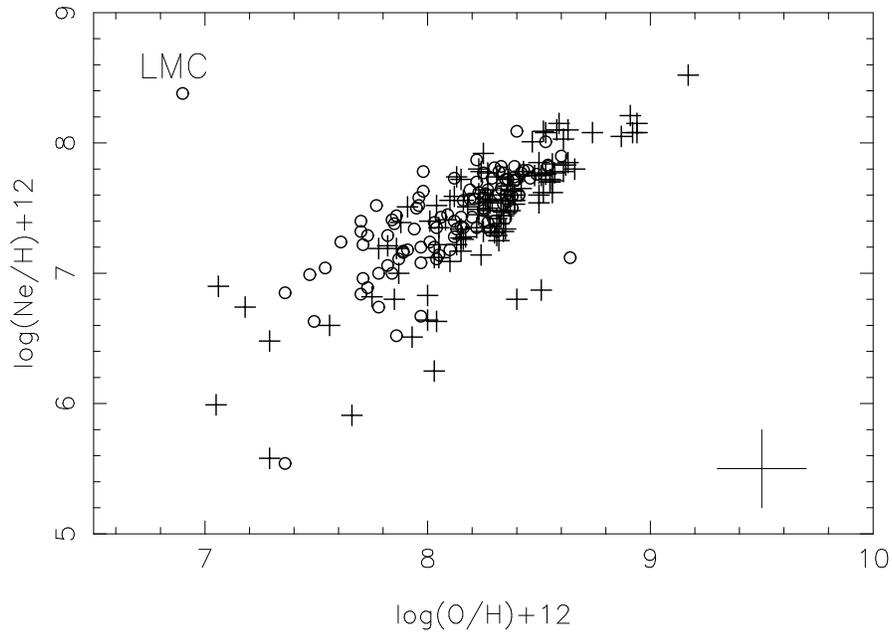}
      \caption{The same as Fig.~8, for the LMC. No IAG/USP data is available for this object.}
   \label{fig9}
   \end{figure}

%oooooooooooooooooooooooooooooooooooooooooooooooooooooooooooooooooooooooooooooooooooooooooooooo
   \begin{figure}
   \centering
   \includegraphics[angle = -90]{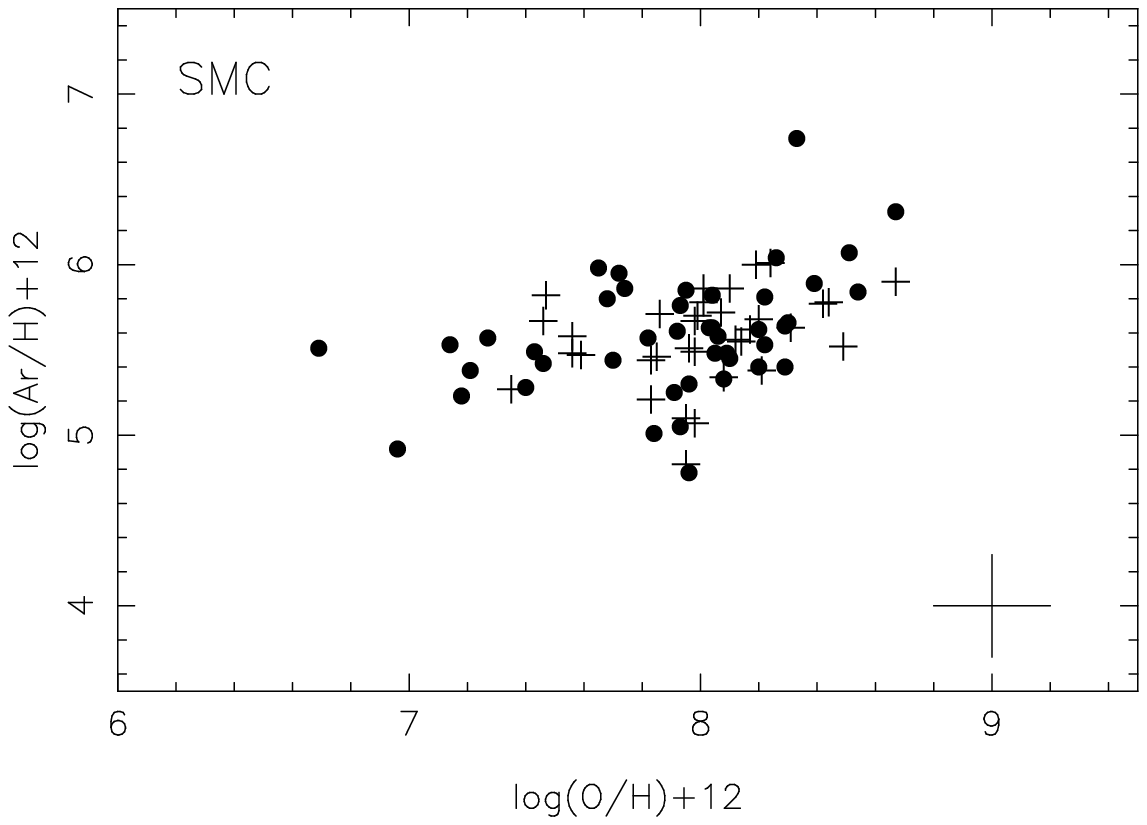}
      \caption{Distance-independent correlation of Ar/H $\times$ O/H for
      the SMC. Symbols are as in Fig.~8.}
   \label{fig10}
   \end{figure}
%oooooooooooooooooooooooooooooooooooooooooooooooooooooooooooooooooooooooooooooooooooooooooooooo
   \begin{figure}
   \centering
   \includegraphics[angle = -90]{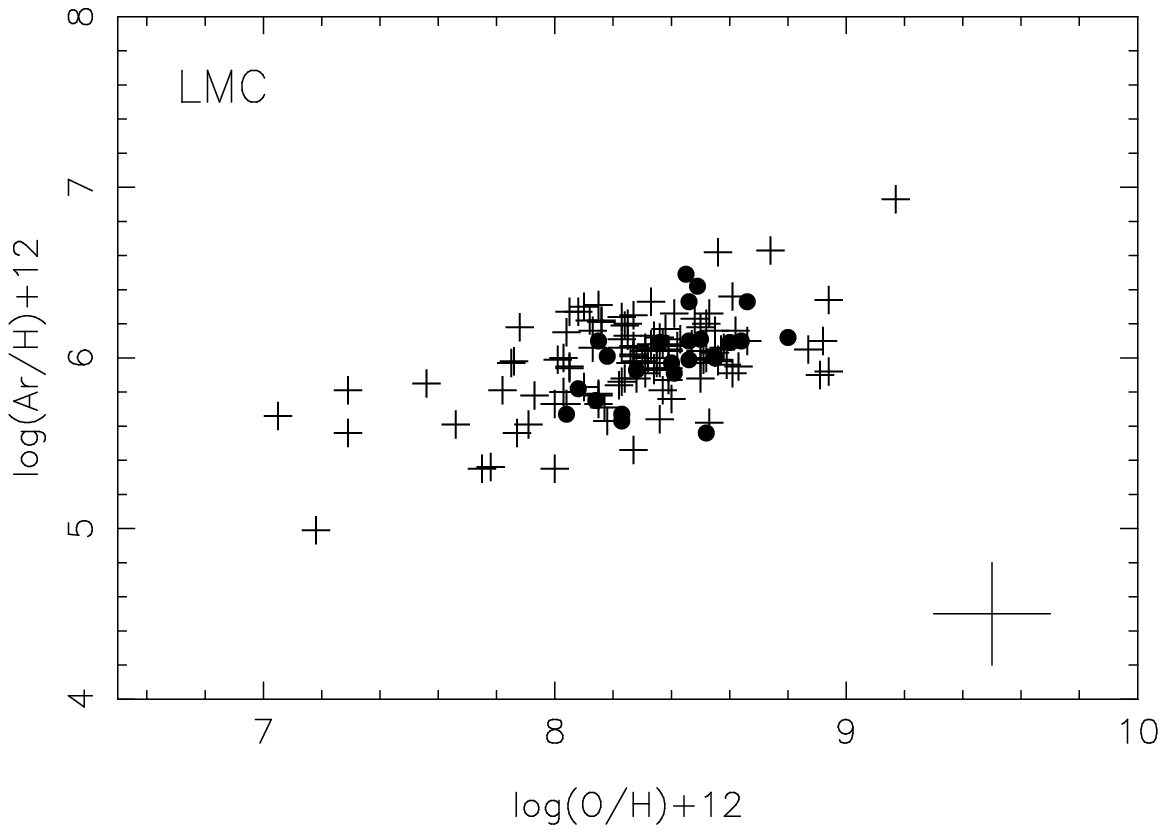}
      \caption{The same as Fig.~10, for the LMC.}
   \label{fig11}
   \end{figure}

%oooooooooooooooooooooooooooooooooooooooooooooooooooooooooooooooooooooooooooooooooooooooooooooo
   \begin{figure}
   \centering
   \includegraphics[angle = -90]{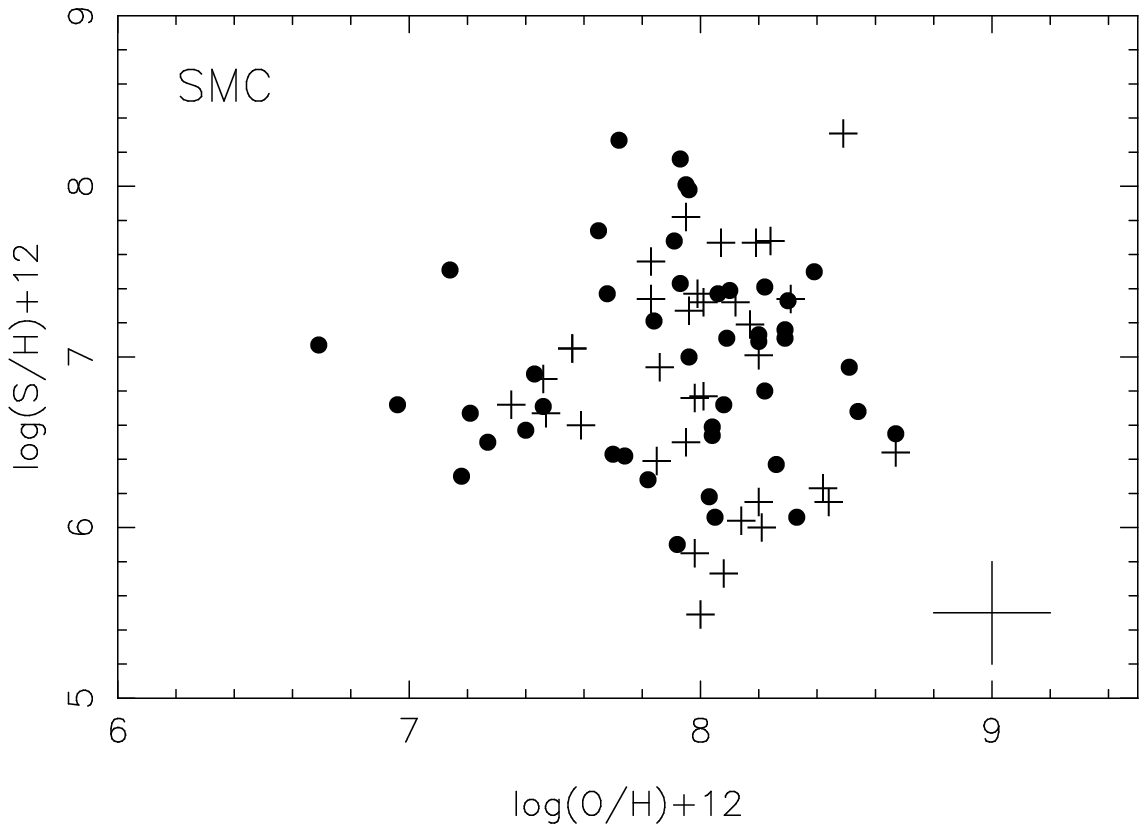}
      \caption{Distance-independent correlation of S/H $\times$ O/H for
      the SMC. Symbols are as in Fig.~8.}
   \label{fig12}
   \end{figure}
%oooooooooooooooooooooooooooooooooooooooooooooooooooooooooooooooooooooooooooooooooooooooooooooo
   \begin{figure}
   \centering
   \includegraphics[angle = -90]{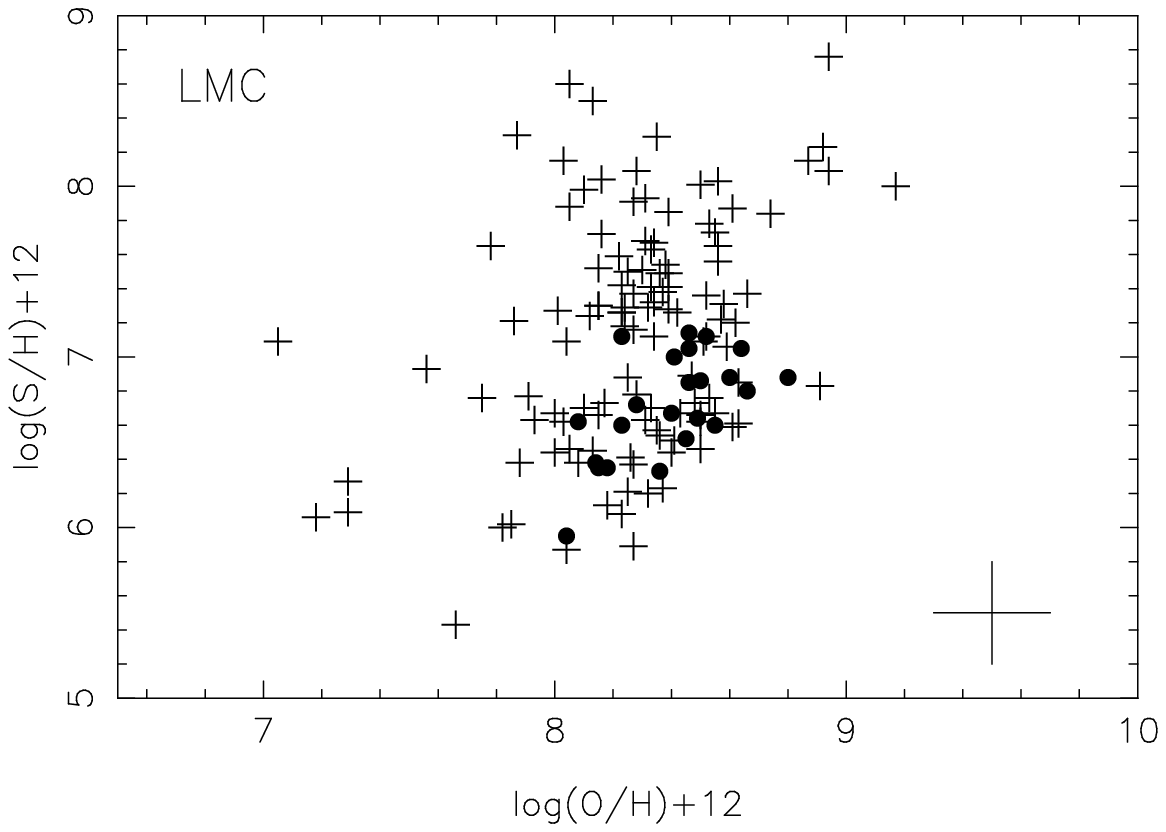}
      \caption{The same as Fig.~12, for the LMC.}
   \label{fig13}
   \end{figure}

%oooooooooooooooooooooooooooooooooooooooooooooooooooooooooooooooooooooooooooooooooooooooooooooo
   \begin{figure}
   \centering
   \includegraphics[angle = -90]{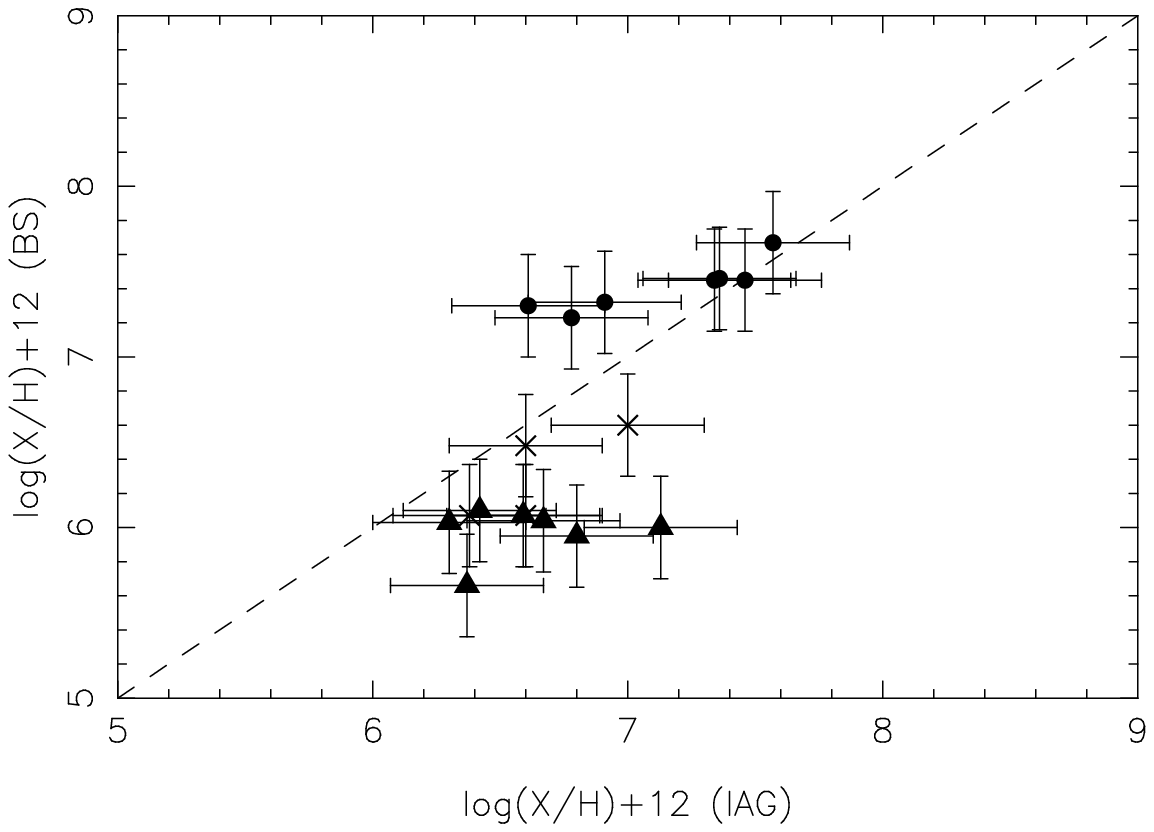}
      \caption{Comparison of the {\it Spitzer} results by Bernard-Salas
       et al. (\citeyear{bernard}) and the IAG/USP sample. circles:
       SMC, Ne/H data; triangles: SMC, S/H abundances; crosses:
       LMC, S/H data.}
   \label{fig14}
   \end{figure}

A hint on the problem of the sulfur abundances in PN can be obtained by comparing our
S/H abundances with the recent determinations by Bernard-Salas et al. (\citeyear{bernard}),
who have presented Ne/H and S/H abundances for 25 PN in the Magellanic Clouds using
{\it Spitzer} data. These results have been obtained on the basis of high-resolution
spectroscopic observations in the infrared, and are in principle more accurate compared
with  the abundances of our present sample, since the uncertainties in the electron
temperatures do not affect the infrared lines, the interstellar extinction effects are
smaller, and the use of the often uncertain ionization correction factors is greatly reduced 
(cf. Bernard-Salas et al. \citeyear{bernard}). A comparison of the Ne/H and S/H abundances 
from this source and those by the IAG/USP group is shown in Fig.~14, where the adopted 
uncertainties are also shown. There are eleven objects in common, which is a small but 
representative sample. In the figure, the circles refer to Ne/H and the triangles for S/H 
for the SMC, while the crosses are S/H data for PN in the LMC.

It can be seen that the Ne/H abundances show a very good agreement with the infrared 
data, while for S/H there is a tedency for our values to be larger than those by
Bernard-Salas et al. (\citeyear{bernard}). Although the differences are not very 
large except for a few nebulae, it may be suggested that the S/H data presented
here should be considered as upper limits. Inspecting Figs.~12 and  13, that would
be expected especially for those nebulae having lower oxygen abundances, which would
explain the scatter diagram observed in Fig.~12. In Bernard-Salas et al.
(\citeyear{bernard}), a similar comparison of the {\it Spitzer} S/H abundances with
data by Leisy \& Dennefeld ({\citeyear{leisy}) was presented, and it was shown that
the latter are also systematically larger than the infrared results. This was 
interpreted by Bernard-Salas et al. (\citeyear{bernard}) as the ionization correction
factors used by Leisy \& Dennefeld (\citeyear{leisy}) overestimated the contribution 
of the S$^{+3}$ ion to the total sulfur abundances. In fact, several of the S/H values 
for Magellanic Cloud PN in Leisy \& Dennefeld (\citeyear{leisy}) are flagged as upper 
limits.  While commenting on the large dispersion of their $\log {\rm S/H} \ \times 
\log {\rm O/H}$ plot, the authors stress  that the sulfur abundances are affected by 
several problems, such as the lack of [SIV]  or [SIII] lines, blending with oxygen lines, 
and innacuracies in the adopted electron temperatures. By considering only the nebulae 
for which the sulfur data is more reliable, Leisy \& Dennefeld (\citeyear{leisy}) obtain 
a somewhat reduced dispersion on the $\log {\rm S/H} \ \times \log {\rm O/H}$ plane, but 
it is still concluded that the sulfur abundances are not good metallicity indicators for 
Magellanic Cloud planetary nebulae.

A discussion of the sulfur abundance problem in PN was recently given by Henry et al. 
(\citeyear{henry04}, \citeyear{henry06}). These authors identified a so-called 
\lq\lq sulfur anomaly\rq\rq, or the lack of agreement of the S/H ratio in PN with
corresponding data from HII regions and other objects. From an analysis of the abundances 
in Milky Way planetary nebulae, HII regions and blue compact galaxies, it was suggested 
that the origin of the \lq\lq sulfur anomaly\rq\rq \ is probably linked to 
the presence of S$^{+3}$ ions, which would affect the total sulfur abundances, at least 
in some nebulae. According to this view, the abundances of at least some of the galactic
PN are underestimated, in the sense that the measured S/H ratio is lower than expected
on the basis of the derived O/H abundances. If this explanation is valid for Magellanic
Cloud PN, it would probably affect those objects with higher O/H ratios, so it is an
alternative to the previous suggestion based on the comparison of optical abundances
with infrared data. 

\noindent
However, other factors may play a role, such as the weakness of the 
sulfur lines, the assumptions leading to the ionization correction factors, etc., so that 
this problem deserves further investigation.

\subsection{Abundance correlations: elements produced by the progenitor stars}

Considering now the elements that are produced during the evolution of the PN progenitor
stars, namely, He and N, Figs.~15 and 16 show the derived correlations of N/H and O/H for
the SMC and LMC, respectively. As expected, a positive correlation is observed, which is 
especially evident in the case of the LMC, but the dispersion of the data is larger  than 
in the case of Ne and Ar. This is due to the fact that the PN display both the original N 
present at the formation of the star and the contamination that is dredged up at the AGB 
branch of the stellar evolution. In other words, the N/H ratio measured in PN shows
some contamination, or enrichment, in comparison with the original abundances in the
progenitor star. 

\bigskip
\bigskip

%oooooooooooooooooooooooooooooooooooooooooooooooooooooooooooooooooooooooooooooooooooooooooooooo
   \begin{figure}
   \centering
   \includegraphics[angle = -90]{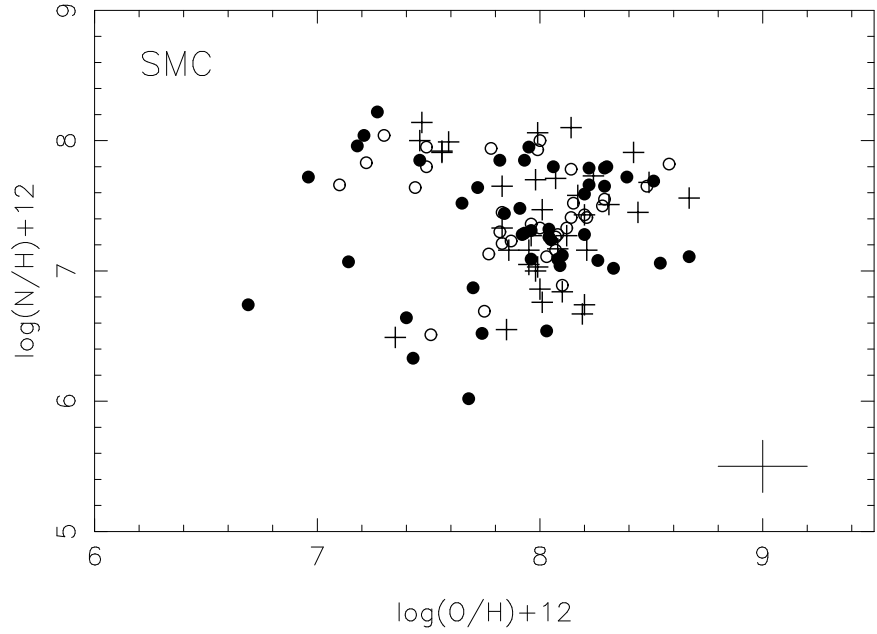}
      \caption{Distance-independent correlation of N/H $\times$ O/H for
      the SMC. Symbols are as in Fig.~8.}
   \label{fig15}
   \end{figure}
%oooooooooooooooooooooooooooooooooooooooooooooooooooooooooooooooooooooooooooooooooooooooooooooo
   \begin{figure}
   \centering
   \includegraphics[angle = -90]{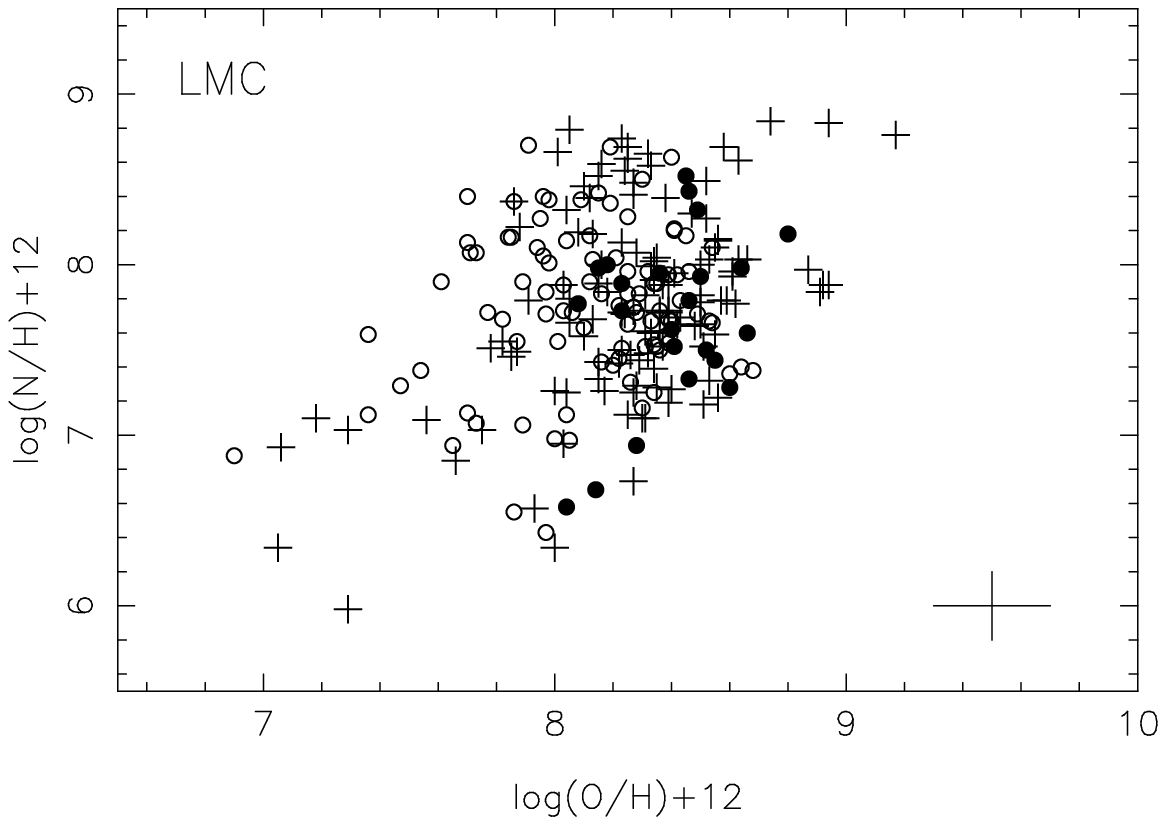}
      \caption{The same as Fig.~15, for the LMC.}
   \label{fig16}
   \end{figure}

%oooooooooooooooooooooooooooooooooooooooooooooooooooooooooooooooooooooooooooooooooooooooooooooo
   \begin{figure}
   \centering
   \includegraphics[angle = -90]{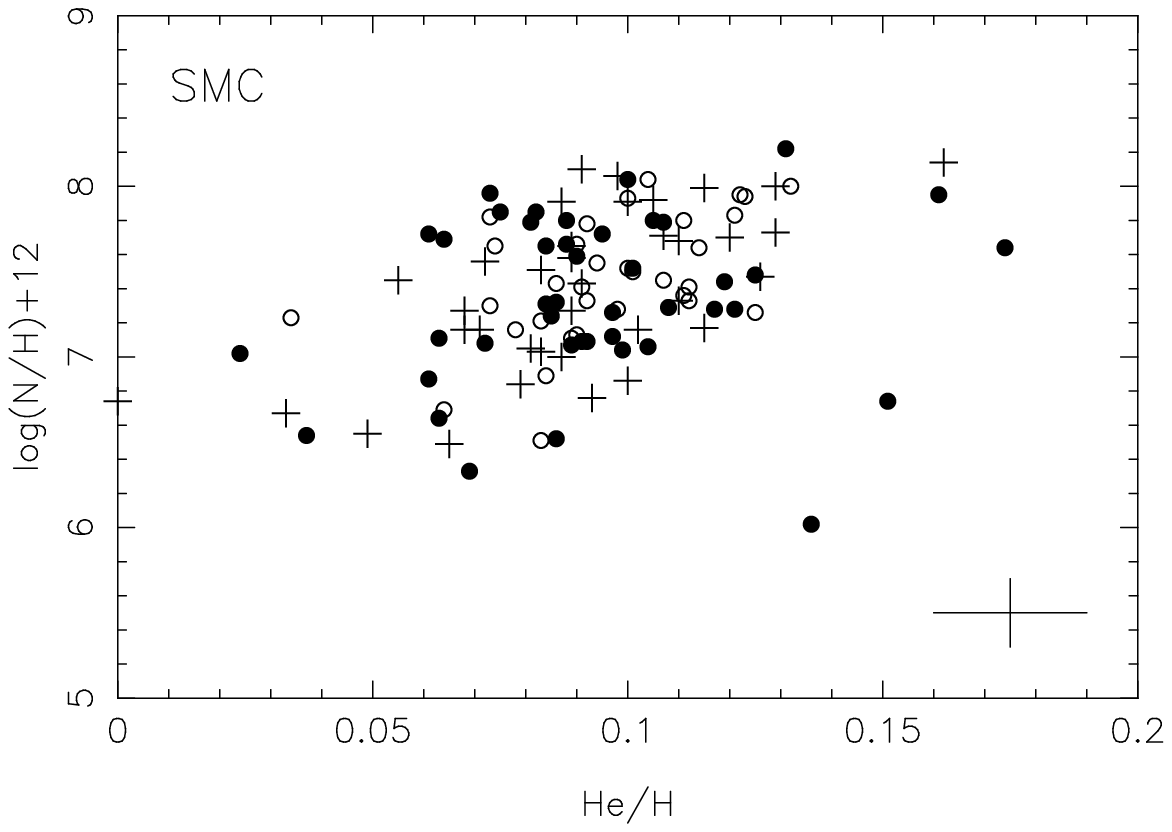}
      \caption{Distance-independent correlation of N/H $\times$ He/H for
      the SMC. Symbols are as in Fig.~8.}
   \label{fig17}
   \end{figure}
%oooooooooooooooooooooooooooooooooooooooooooooooooooooooooooooooooooooooooooooooooooooooooooooo
   \begin{figure}
   \centering
   \includegraphics[angle = -90]{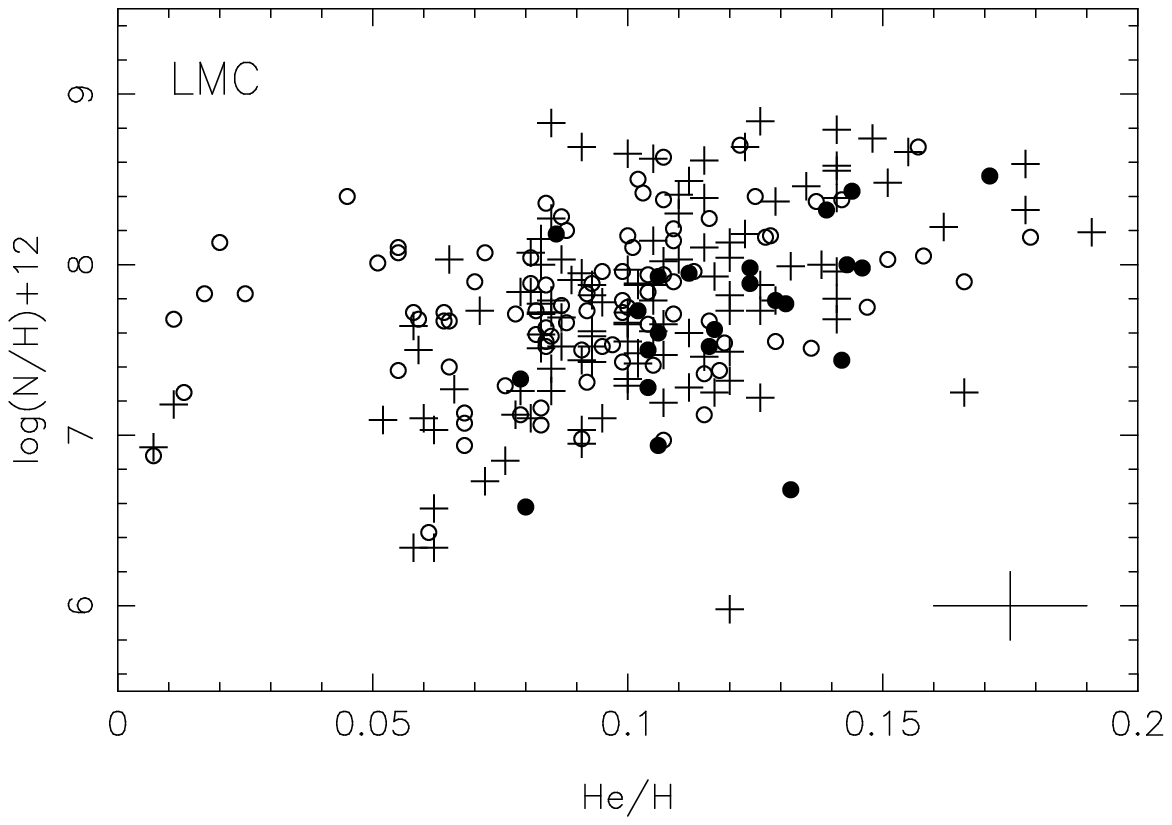}
      \caption{The same as Fig.~17, for the LMC.}
   \label{fig18}
   \end{figure}

%oooooooooooooooooooooooooooooooooooooooooooooooooooooooooooooooooooooooooooooooooooooooooooooo
   \begin{figure}
   \centering
   \includegraphics[angle = -90]{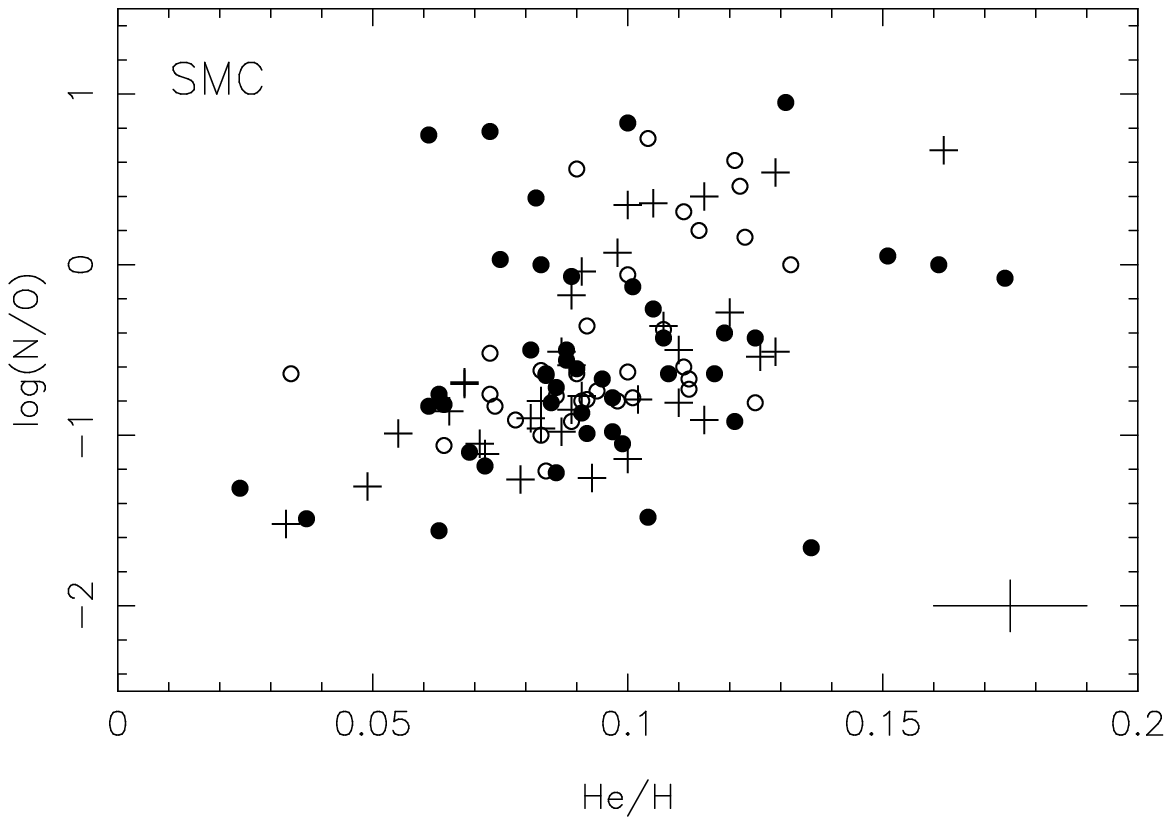}
      \caption{Distance-independent correlation of N/O $\times$ He/H for
      the SMC. Symbols are as in Fig.~8.}
   \label{fig19}
   \end{figure}
%oooooooooooooooooooooooooooooooooooooooooooooooooooooooooooooooooooooooooooooooooooooooooooooo
   \begin{figure}
   \centering
   \includegraphics[angle = -90]{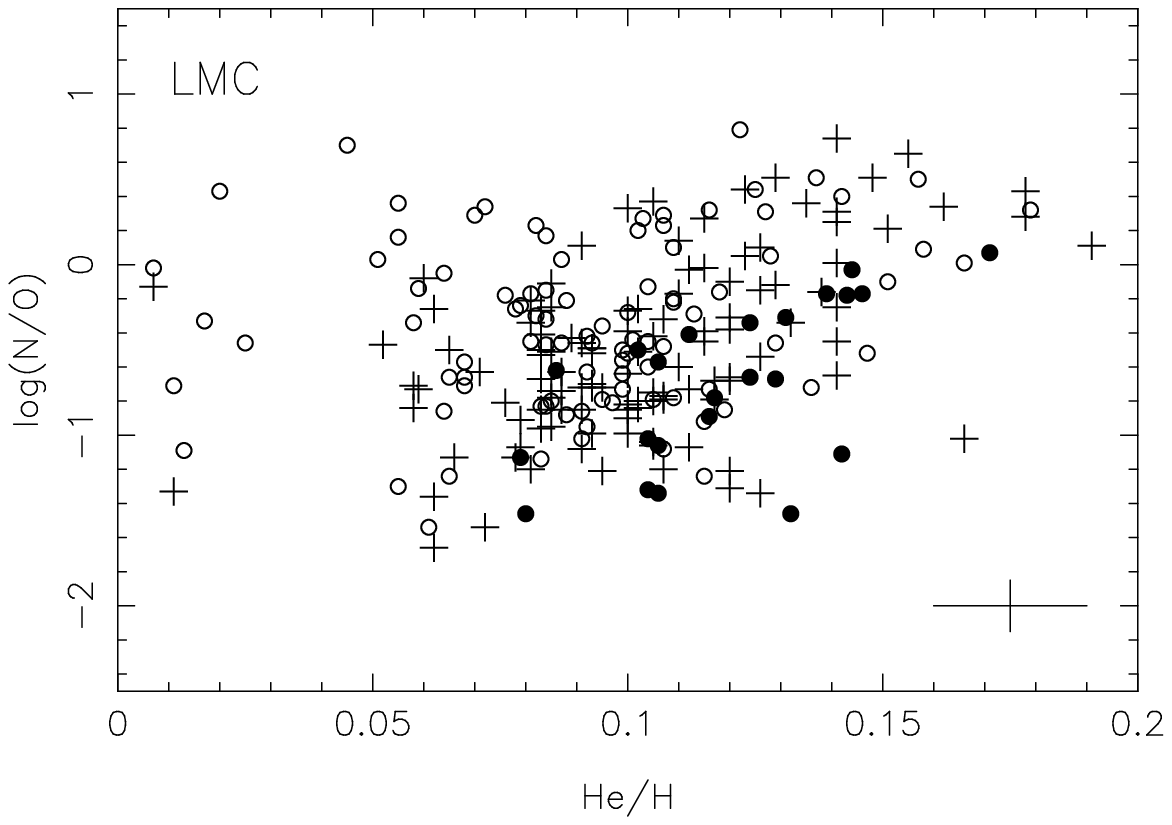}
      \caption{The same as Fig.~19, for the LMC.}
   \label{fig20}
   \end{figure}

%oooooooooooooooooooooooooooooooooooooooooooooooooooooooooooooooooooooooooooooooooooooooooooooo
   \begin{figure}
   \centering
   \includegraphics[angle=-90]{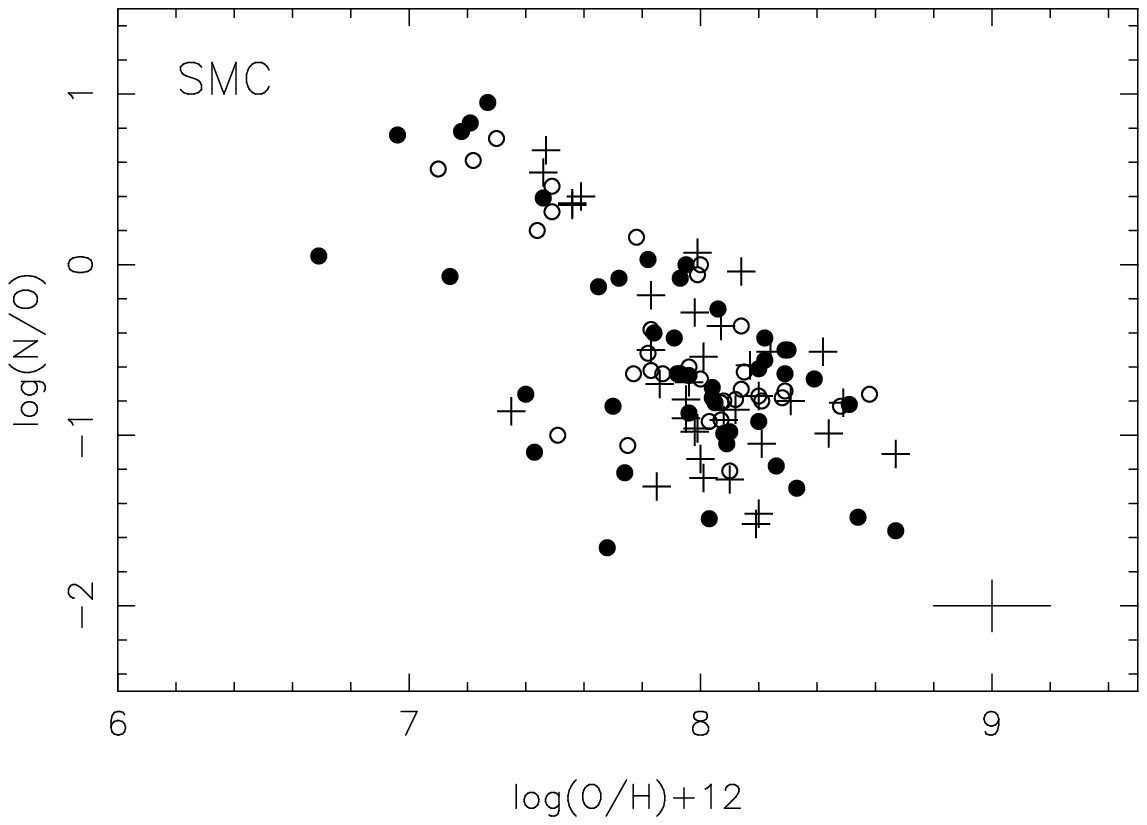}
      \caption{Distance-independent correlation of N/O $\times$ O/H for
      the SMC. Symbols are as in Fig.~8.}
   \label{fig21}
   \end{figure}
%oooooooooooooooooooooooooooooooooooooooooooooooooooooooooooooooooooooooooooooooooooooooooooooo
   \begin{figure}
   \centering
   \includegraphics[angle = -90]{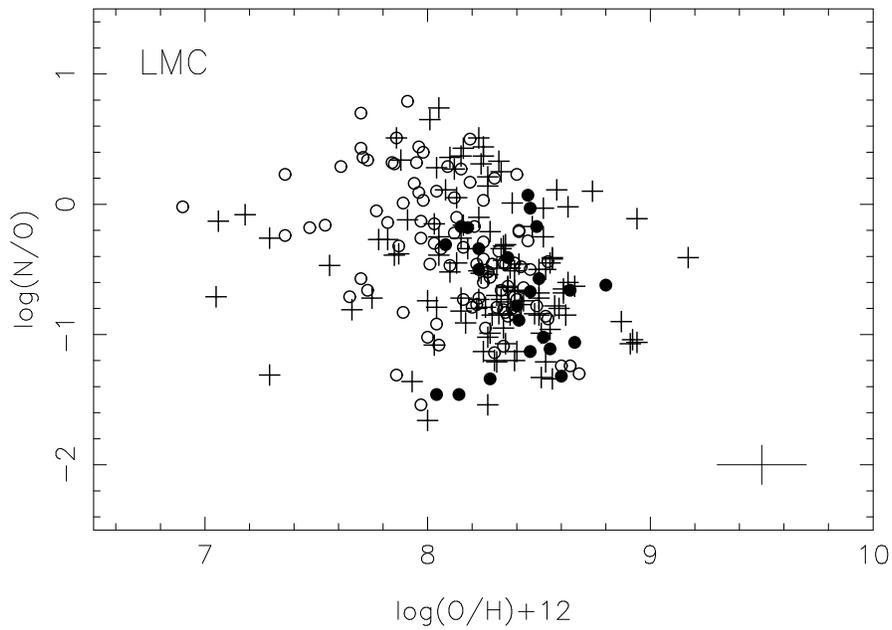}
      \caption{The same as Fig.~21, for the LMC.}
   \label{fig22}
   \end{figure}

An estimate of the nitrogen enrichment from the PN 
progenitor stars can be made by comparing the average N/H abundances of Table~1 with
those of HII regions. The Orion value given in the table is similar to the PN abundances
for the 3 samples considered, but HII regions in the lower metallicity Magellanic Clouds
have accordingly lower nitrogen abundances. As an example, for 30 Doradus, the brightest
HII region in the LMC, Peimbert (\citeyear{peimbert}) estimates $\epsilon({\rm N}) =
7.05$ based on echelle spectrophotometry, assuming no temperature fluctuations ($t^2 = 0.00$). 
Comparing this result with the data of Table~1, an average enrichment of about 0.6--0.7 dex 
is obtained for the N/H ratio. Concerning HII regions in the SMC, Rela\~no et al. 
(\citeyear{relano}) estimate $\epsilon({\rm N}) = 6.81$ for NGC~346 on the basis of 
photoionization models, which implies an enrichment of 0.5--0.6 dex for the PN samples 
listed in Table~1. These  enrichment factors may be affected by the chemical evolution of 
the host galaxy, which includes the average  increase of the metallicity as the galaxy evolves, 
but it is interesting that similar factors are obtained both for the LMC and SMC. The quoted 
values for 30 Dor and NGC~346 are included at the bottom of Table~1, as representative 
of HII in the Magellanic Clouds.

Figs.~17 and 18 show the N/H abundances as a function of the He/H ratio, while Figs.~19 and
20 are the corresponding plots for N/O as a function of He/H. As pointed
out in the literature (cf. Kwok \citeyear{kwok}), these ratios present enhancements
relative to the average interstellar values. The dispersion is again large, but a positive 
correlation can also be observed, as expected, since the same processes that increase
the nitrogen abundances in PN also affect the He/H ratio. A plot similar to Figs.~19 and 20
was presented by Shaw (\citeyear{shaw}), in an effort to separate PN of different 
morphologies. In the LMC, some objects in the  sample by Stasi\'nska et al. (\citeyear{stasinska}) 
have very low He abundances while the N/O ratio is normal, suggesting that neutral helium may be 
present in these objects. As pointed out by Maciel et al. (\citeyear{maciel2006a}), the N/O 
$\times$ \ He/H ratios in the Magellanic Clouds support the correlation observed in the Milky 
Way, but the N/O ratio is comparatively lower. The O/H ratio corresponding to the SMC is also 
lower, which can be interpreted as an evidence that the lower metallicity environment in the 
SMC leads to a smaller fraction of Type I PN, which are formed by the more massive stars in 
the Intermediate Mass Star bracket (cf. Stanghellini et al. \citeyear{stanghellini03}).

Finally, Figs.~21 and 22 show the N/O ratios as a function of the O/H abundances. The 
conversion of oxygen into nitrogen by the ON cycling in the PN progenitor stars has been 
suggested in the literature as an explanation for the anticorrelation between N/O and
O/H in planetary nebulae (cf. Costa et al. \citeyear{costa1}, Stasi\'nska et al.
\citeyear{stasinska}, Perinotto et al. \citeyear{perinotto}). This relation is approximately
valid on the basis of PN data in several galaxies of the Local Group, as discussed by
Richer \& McCall (\citeyear{richer}). From Figs.~21 and 22, we conclude that the Magellanic
Cloud data support such anticorrelation, particularly in the case of the SMC. As discussed
by Maciel et al. (\citeyear{maciel2006a}) the Milky Way data define a mild anticorrelation,
especially in the case of $\epsilon({\rm O}) = \log ({\rm O/H}) + 12 > 8.0$, which is better defined 
by the SMC/LMC.

\bigskip
{\it Acknowledgements. This work was partly supported by FAPESP and CNPq.}

\end{document}